\pdfoutput=1

\documentclass[11pt,twoside,a4paper,cmspaper,final,collab]{cms-tdr}
\def\svnVersion{438055P}\def\cmsCernNoTag{CERN-EP-2017-077}\def\cmsCernDate{\today}\def\cmsMessage{Published in Physics Letters B as \href{http://dx.doi.org/10.1016/j.physletb.2017.09.083}{\doi{10.1016/j.physletb.2017.09.083.}}}

\begin{document}\cmsNoteHeader{B2G-16-007}

\hyphenation{had-ron-i-za-tion}
\hyphenation{cal-or-i-me-ter}
\hyphenation{de-vices}
\RCS$Revision: 438044 $
\RCS$HeadURL: svn+ssh://svn.cern.ch/reps/tdr2/papers/B2G-16-007/trunk/B2G-16-007.tex $
\RCS$Id: B2G-16-007.tex 438044 2017-12-07 18:10:42Z jngadiub $
\newlength\cmsFigWidth
\ifthenelse{\boolean{cms@external}}{\setlength\cmsFigWidth{0.85\columnwidth}}{\setlength\cmsFigWidth{0.4\textwidth}}
\ifthenelse{\boolean{cms@external}}{\providecommand{\cmsLeft}{top\xspace}}{\providecommand{\cmsLeft}{left\xspace}}
\ifthenelse{\boolean{cms@external}}{\providecommand{\cmsRight}{bottom\xspace}}{\providecommand{\cmsRight}{right\xspace}}
\ifthenelse{\boolean{cms@external}}{\providecommand{\cmsTableResize}[1]{#1}}{\providecommand{\cmsTableResize}[1]{\resizebox{\textwidth}{!}{#1}}}
\newcommand{\mJ}{\ensuremath{m_{\text{jet}}}\xspace}
\newcommand{\qqbarpr}{\ensuremath{\Pq\Paq^{(\prime)}}\xspace}
\newcommand{\nsubj}{\ensuremath{\tau_{21}}}
\newcommand{\Gbulk}{\ensuremath{\mathrm{G}_\text{bulk}}\xspace}
\newcommand{\Vprime}{\ensuremath{\mathrm{V}'}\xspace}
\providecommand{\NA}{\ensuremath{\text{---}}\xspace}

\cmsNoteHeader{B2G-16-007}
\title{Combination of searches for heavy resonances decaying to WW, WZ, ZZ, WH, and ZH boson pairs in proton-proton collisions at $\sqrt{s} = 8$ and 13\TeV}

\author[unesp]	{	Sudha	Ahuja	}
\author[ttu]    {   Nural	Akchurin	}
\author[uzh]	{	Thea Klaeboe Aarrestad	}
\author[milano]	{	Luca	Brianza	}
\author[ncu]	{	Yu-Hsiang Chang	}
\author[ncu]	{	Ching-Wei Chen	}
\author[ttu]	{	Jordan	Damgov	}
\author[ttu]	{	Phil	Dudero	}
\author[ulb]    {       Laurent Favart  }
\author[milano] {       Raffaele Gerosa }
\author[milano] {       Alessio Ghezzi  }
\author[lyon]   {       Maxime  Gouzevitch      }
\author[milano] {       Pietro  Govoni  }
\author[fnal]	{	Lindsey Gray 	}
\author[uzh]	{	Andreas	Hinzmann	}
\author[pku]    {       Huang   Huang     }
\author[ncu]	{	Ji-Kong	Huang	}
\author[ncu]	{	Raman	Khurana	}
\author[uzh]	{	Ben	Kilminster	}
\author[uzh]	{	Clemens	Lange	}
\author[ttu]	{	Sung-Won	Lee	}
\author[pku]	{	Qiang	Li	}
\author[ncu]	{	Yun-Ju	Lu	}
\author[jhu]	{	Petar	Maksimovic	}
\author[uzh]	{	Jennifer	Ngadiuba	}
\author[unesp]	{	Sergio	Novaes	}
\author[padova]	{	Alexandra	Oliveira	}
\author[padova]	{	Alberto Zucchetta	}
\author[padova]	{	Jacopo Pazzini	}
\author[caltech] { Maurizio Pierini}
\author[buffalo]	{	Salvatore	Rappoccio	}
\author[unesp]  {       Jos\'e Ruiz     }
\author[unesp]	{	Thiago	Tomei	}
\author[ncu]	{	Henry Yee-Shian	Tong	}
\author[fnal]	{	Nhan	Tran	}
\author[pku]	{	Qun	Wang	}
\author[pku]	{	Mengmeng	Wang	}
\author[ncu]	{	Jun-Yi	Wu	}
\author[pku]	{	Zijun	Xu	}
\author[ncu]	{	Shin-Shan Eiko	Yu	}
\author[pku]    {       Xiaoqing Yuan      }

\date{\today}

\abstract{
A statistical combination of searches is presented for massive resonances decaying to WW, WZ, ZZ, WH, and ZH boson pairs in proton-proton collision data collected by the CMS experiment at the LHC.
The data were taken at centre-of-mass energies of 8 and 13\TeV, corresponding to respective integrated luminosities of 19.7 and up to 2.7\fbinv.
The results are interpreted in the context of heavy vector triplet and singlet models that mimic properties of composite-Higgs models predicting \PWpr and \PZpr bosons decaying to WZ, WW, WH, and ZH bosons.
A model with a bulk graviton that decays into WW and ZZ is also considered.
This is the first combined search for WW, WZ, WH, and ZH resonances and yields lower limits on masses at 95\% confidence level for \PWpr and \PZpr singlets at 2.3\TeV, and for a triplet at 2.4\TeV.
The limits on the production cross section of a narrow bulk graviton resonance with the curvature scale of the warped extra dimension $\tilde{k}=0.5$, in the mass range of 0.6 to 4.0\TeV, are the most stringent published to date.
}

\hypersetup{%
pdfauthor={CMS Collaboration},%
pdftitle={Combination of searches for heavy resonances decaying to WW, WZ, ZZ, WH, and ZH boson pairs in proton-proton collisions at sqrt(s) = 8 and 13 TeV},%
pdfsubject={CMS},%
pdfkeywords={CMS, physics, di-boson, resonances, combination}}

\maketitle
\section{Introduction}
\label{sec:introduction}
Hypotheses for physics beyond the standard model (SM) predict the existence of heavy resonances that decay to any combination of two among the massive vector bosons (W or Z, collectively referred to as V) or to a V and the scalar SM Higgs boson (H). Among the considered models are those dealing with warped extra dimensions (WED)~\cite{Randall:1999ee,Randall:1999vf} and composite-Higgs bosons~\cite{Bellazzini:2014yua,CHM2,Composite2,Greco:2014aza}.
Searches for such VV and VH resonances in different final states have previously been performed by the ATLAS~\cite{Aad:2015ipg,Aad:2015ufa,ATLASwprimeWZPAS,Aaboud:2016okv, Aaboud:2016lwx,Aad:2015yza} and CMS~\cite{Khachatryan:2014xja,Khachatryan:2014gha,Khachatryan:2014hpa,Khachatryan:2016yji,Khachatryan:2015ywa,Khachatryan:2015bma,Sirunyan:2016cao,Khachatryan:2016cfx} experiments at the CERN LHC.
As all of these searches have similar sensitivities, a statistical combination of the CMS results is provided to improve the overall result.
The current status of heavy diboson searches at CMS is also of interest in this respect, with recent work in the all-jet VV~\cite{Aad:2015owa} and lepton+jet WH~\cite{Khachatryan:2016yji} decay channels showing possible enhancements.

The benchmark models considered in combining the results are a heavy vector triplet (HVT) model~\cite{Pappadopulo:2014qza} and the bulk scenario~\cite{Agashe:2007zd, Fitzpatrick:2007qr, Antipin:2007pi} ($\Gbulk$ graviton) in the Randall--Sundrum (RS) WED model~\cite{Randall:1999ee,Randall:1999vf}.
The HVT model generalizes a large number of models that predict spin-1 resonances, such as those in composite-Higgs theories, which can arise as a singlet, either \PWpr or \PZpr~\cite{Grojean:2011vu,Langacker:2008yv,Salvioni:2009mt}, or as a $\Vprime$ triplet (where $\Vprime$ represents $\PWpr$ and $\PZpr$ bosons)~\cite{Pappadopulo:2014qza}.
The HVT and $\Gbulk$ models are considered as benchmarks for diboson resonances with spin~1 ($\PWpr\to\PW\Z$ or WH, $\PZpr\to\PW\PW$ or ZH), and spin~2 ($\Gbulk\to\PW\PW$ or ZZ), respectively, produced via quark-antiquark annihilation ($\qqbar'\to\PWpr$, $\qqbar\to\PZpr$) and gluon-gluon fusion ($\Pg\Pg\to \Gbulk$).

The analyses included in this statistical combination are based on proton-proton (pp) collision data collected by the CMS experiment~\cite{Chatrchyan:2008zzk} at $\sqrt{s} = 8$ and 13\TeV, corresponding to respective integrated luminosities of 19.7 and 2.3--2.7\fbinv.
Of the 2.7\fbinv recorded at 13\TeV, the detector was fully operational for 2.3\fbinv, while 0.4\fbinv were collected with only the central  part of the detector ($\abs{\eta}<3$) in optimal condition.
The signal corresponds to a narrow charge 0 or 1 resonance with a mass $>$0.6\TeV that decays to any of the two high energy W, Z, or Higgs bosons, where narrow refers to the assumption that the natural relative width is smaller than the typical experimental resolution of 5\%, which is true for a large fraction of the parameter space of the reference models.
For the mass range under study, the particles emerging from the boson decays are highly collimated, requiring special reconstruction and identification techniques that are in common in these kinds of analyses.

Analyses were performed using all-lepton, lepton+jet, and all-jet final states that include decays of W and Z bosons into charged leptons ($\ell=\Pe$ or $\PGm$) and neutrinos ($\Pgn$), as well as the reconstructed jets evolved from the \qqbarpr products of the boson decays. The latter include $\PW\to\qqbar'$ and $\PZ\to\qqbar$.
The analyses use $\PH\to\bbbar$ and $\PH\to\PW\PW\to\qqbar'\qqbar'$ decays of the Higgs boson, which are labeled as \bbbar or $\qqbar\qqbar$, together with a vector boson decaying to hadrons.
Final states with the Higgs boson decaying into a $\tau^+\tau^-$ lepton pair are also considered.
In all, we combine results from the following final states:
3$\ell\Pgn$ (8\TeV)~\cite{Khachatryan:2014xja};
$\ell\ell \qqbar$ (8\TeV)~\cite{Khachatryan:2014gha};
$\ell \Pgn \qqbar$ (8\TeV)~\cite{Khachatryan:2014gha};
$\qqbar \qqbar$ (8\TeV)~\cite{Khachatryan:2014hpa};
$\ell \Pgn \bbbar$ (8\TeV)~\cite{Khachatryan:2016yji};
$\qqbar \tau\tau $ (8\TeV)~\cite{Khachatryan:2015ywa};
$\qqbar\bbbar$ and $6\Pq$ (8\TeV)~\cite{Khachatryan:2015bma};
$\ell \Pgn \qqbar$ (13\TeV)~\cite{Sirunyan:2016cao};
$\qqbar \qqbar$ (13\TeV)~\cite{Sirunyan:2016cao};
and $\ell \ell \bbbar$, $\ell \Pgn \bbbar$, and $\Pgn \Pgn \bbbar$ (13\TeV)~\cite{Khachatryan:2016cfx}.
Since some more forward parts of the detector, which provide information for the calculation of the missing transverse momentum, were not in optimal condition for a fraction of the 2015 data-taking period, the analyses of 13\TeV data in the $\ell\nu\qqbar$, $\ell\nu\bbbar$, $\ell\ell\bbbar$, and $\nu\nu\bbbar$ decay channels are based on the dataset corresponding to the integrated luminosity of 2.3\fbinv rather than 2.7\fbinv.

Given the limited experimental jet mass resolution, the $\PW\to\qqbar'$ and $\PZ\to\qqbar$ candidates cannot be fully differentiated, and individual analyses can be sensitive to several different interpretations in the same model.
For example, the final state $\ell \Pgn \qqbar$ is sensitive to HVT \PWpr decays to a WZ boson pair as well as to \PZpr decays to WW boson pairs.
The sum of contributions from multiple signals with their respective efficiencies is sought in the combination.
For this reason, separate interpretations are given below for a vector triplet $\Vprime$ and for vector singlets (\PWpr or \PZpr).

This letter is structured as follows.
After a brief introduction to the benchmark models in Section~\ref{sec:models}, a summary of the analyses entering the combination is given in Section~\ref{sec:analyses}.
The combining procedure is described in Section~\ref{sec:combination}, and finally the results and summary are provided in Sections~\ref{sec:results} and~\ref{sec:conclusions}.

\section{Theoretical models}
\label{sec:models}
As indicated above, heavy diboson resonances are expected in a large class of models that attempt to accommodate the difference between the electroweak and Planck scales.
We perform the combination in the context of seven benchmark theories formulated to cover different spin, production, and decay options for resonances decaying to VV and VH.
The properties of models for spin-1 and spin-2 resonances are briefly discussed in the following two subsections, with benchmark resonances summarized in Table~\ref{tab:models}. For both spin-1 and spin-2 resonances, the signal cross sections used in this paper are given in Tables~\ref{tab:xsec8} and~\ref{tab:xsec13} of the Appendix.

\subsection{Spin-1 resonances}
\label{subsec:hvt}

Several extensions of the SM such as composite-Higgs \cite{Bellazzini:2014yua,CHM2,Composite2,Greco:2014aza} and little Higgs \cite{Schmaltz:2005ky,ArkaniHamed:2002qy} models
can be generalized through a phenomenological Lagrangian that describes the production and decay of spin-1 heavy resonances, such as a charged \PWpr{} and a neutral \cPZpr{}, using the HVT model.

The HVT couplings are described in terms of four parameters:
\begin{enumerate}
\item[(i)] $c_\mathrm{H}$ describes interactions of the new resonance with the Higgs boson or longitudinally polarized SM vector bosons;
\item[(ii)] $c_\mathrm{F}$ describes the interactions of the new resonance with fermions;
\item[(iii)] $g_\mathrm{V}$ gives the typical strength of the new interaction and
\item[(iv)] $m_\mathrm{V}'$ is the mass of the new resonance.
\end{enumerate}
The \PWpr{} and \cPZpr{} bosons couple to the fermions through the combination of parameters $g^2 c_\mathrm{F}/ g_\mathrm{V}$ and to the H and vector bosons through $g_\mathrm{V} c_\mathrm{H}$ , where $g$ is the SU(2)$_L$ gauge coupling.
The Higgs boson is assumed to be part of a Higgs doublet field. Therefore, its dynamics are related to the Goldstone bosons in the same doublet by SM symmetry.
Those Goldstone bosons are equivalent to the corresponding longitudinally polarised W and Z bosons in the high energy limit according to the ``Equivalence Theorem''~\cite{Chanowitz:1985hj}. The coupling of the Higgs boson to the \PWpr and \PZpr resonances can thus be described by the same coupling as used for the longitudinal W and Z bosons.

The production of \PWpr{} and \cPZpr{} bosons at hadron colliders is expected to be dominated by the process $\qqbarpr \to \PWpr{}$ or $\cPZpr{}$.
Two benchmark models are studied, denoted A and B, that were suggested in Ref.~\cite{Pappadopulo:2014qza}.
In model A, weakly coupled vector resonances arise from an extension of the SM gauge group.
In model B, the heavy vector triplet is produced by a strong coupling mechanism, as embodied in theories such as in the composite-Higgs model.
Consequently, in model A the branching fractions to fermions and SM massive bosons are comparable, whereas in model B, fermionic couplings are suppressed.
Therefore, in the context of WW, WZ, ZH, and WH resonance searches, model B is of more interest, since model A is strongly constrained by searches in final states with fermions.
In both options, the heavy resonances couple as SM custodial triplets, so that \PWpr{} and \PZpr{} are expected to be approximately degenerate in mass, and the branching fractions $\mathcal{B}(\PWpr \to \PW\PH)$ and $\mathcal{B}(\PZpr \to \Z\PH)$ to be comparable to $\mathcal{B}(\PWpr \to \PW\Z)$ and $\mathcal{B}(\PZpr\to\PW\PW)$.
We consider model A ($c_\mathrm{H} = -g^2/g_\mathrm{V}^2$, $c_\mathrm{F} = -1.3$) with parameter $g_\mathrm{V} = 1$, and model B ($c_\mathrm{H} = -1$, $c_\mathrm{F} = 1$) with parameter $g_\mathrm{V} = 3$.
A value of $g_\mathrm{V} = 3$ is chosen for model B to represent strongly coupled electroweak symmetry breaking, e.g. composite-Higgs models, while assuring small natural widths relative to the experimental resolution.
We also consider heavy resonances that couple to \PWpr{} and \cPZpr{} as singlets, i.e. expecting only one charged or neutral resonance at a given mass, as summarized in Table~\ref{tab:models}.

Previous searches for a \PWpr boson decaying into a pair of SM massive bosons (WZ, WH) provide a lower mass limit of 1.8\TeV in model A ($g_\mathrm{V} = 1$) and 2.3\TeV in model B ($g_\mathrm{V} = 3$), where the results from 8\TeV data~\cite{Aad:2015ipg,Khachatryan:2014hpa,Khachatryan:2014xja,Khachatryan:2016yji,Aad:2015ufa,ATLASwprimeWZPAS} are most stringent at low resonance masses, while \unit{13}{\TeV} analyses~\cite{Sirunyan:2016cao,Khachatryan:2016cfx, Aaboud:2016okv, Aaboud:2016lwx} dominate at higher resonance masses.
Searches for a \PZpr boson decaying into a pair of SM massive bosons (WW, ZH) yield lower mass limits of 1.4 and 2.0\TeV in models A and B, respectively, based on 8\TeV~\cite{Aad:2015yza,Khachatryan:2015bma,Khachatryan:2015ywa} and \unit{13}{\TeV}~\cite{Sirunyan:2016cao,Khachatryan:2016cfx, Aaboud:2016okv, Aaboud:2016lwx} data.
For a heavy vector triplet resonance, the most stringent lower mass limits of 2.35\TeV (model A) and 2.60\TeV (model B) are obtained from a combination of VV searches at 13\TeV~\cite{Aaboud:2016okv}.

\subsection{Spin-2 resonances}

Massive spin-2 resonances can be motivated in WED models through Kaluza--Klein (KK) gravitons~\cite{Randall:1999ee,Randall:1999vf},
which correspond to a tower of KK excitations of a spin-2 graviton.
The original RS model (here denoted as RS1) can be extended to
the bulk scenario ($\Gbulk$), which addresses the flavor structure of
the SM through the localization of fermions in the warped extra
dimension~\cite{Agashe:2007zd, Fitzpatrick:2007qr, Antipin:2007pi}.

These WED models have two free parameters: the mass of the first mode of the
KK graviton, $m_G$, and the ratio $\tilde{k} \equiv k/\overline{m}_{\text{Pl}}$, where $k$ is the curvature scale of the WED
and $\overline{m}_{\text{Pl}} \equiv m_{\text{Pl}} / \sqrt{8\pi}$ is the reduced Planck mass.
The constant $\tilde{k}$ acts as the coupling constant of the model, on which the production cross sections and widths of the graviton depend quadratically.
For models with $\tilde{k} \lesssim 0.5$, the natural width of the resonance is sufficiently small to be neglected relative to detector resolution.

In the bulk scenario, coupling of the graviton to light fermions is highly
suppressed, and the decay into photons is negligible,
while in the RS1 scenario, the graviton decays to photon and fermion pairs dominate.
In the context of WW and ZZ resonance searches, the bulk scenario is of great interest, since RS1 is already strongly constrained through searches in final states with fermions and photons~\cite{Sirunyan:2016iap,Khachatryan:2016zqb,Khachatryan:2016yec}.
The production of gravitons at hadron colliders in the bulk scenario is dominated by gluon-gluon fusion, and the branching fraction $\mathcal{B}(\Gbulk \to \PW\PW) \approx 2 \, \mathcal{B}(\Gbulk \to \PZ\PZ)$.
The decay mode into a pair of Higgs bosons, which is not studied in this paper, has a branching fraction comparable to $\mathcal{B}(\Gbulk \to \PZ\PZ)$.

For $\tilde{k}=1$, where the bulk graviton has comparable or larger width than the detector resolution, the most stringent lower limit of 1.1\TeV on its mass is set by a combination of searches in the diboson final state~\cite{Aaboud:2016okv}.
The most stringent limits on the cross section for narrow bulk graviton resonances for $\tilde{k} \leq 0.5$ are also determined through searches in the diboson final state~\cite{Sirunyan:2016cao,Khachatryan:2014gha,Khachatryan:2014gha,Khachatryan:2014hpa}; however, the integrated luminosity of the dataset is not large enough to allow us to obtain mass limits for this resonance.

\begin{table*}[htb]
\centering
\topcaption{Summary of the properties of the heavy-resonance models considered in the combination. The polarization of the produced W and Z bosons in these models is primarily longitudinal, as decays to transverse polarizations are suppressed.}
\renewcommand{\arraystretch}{1.2}
\cmsTableResize{
\begin{tabular}{*{6}{c}}
Model & Particles & Spin & Charge & Main production mode & Main decay mode \\
\hline
\multirow{3}{*}{HVT model A, $g_\mathrm{V} = 1$} & \PWpr singlet & 1 & $\pm$1 & $\qqbar'$ & $\qqbar'$ \\
                                                                                 & \PZpr singlet & 1 & 0 & $\qqbar$ &  $\qqbar$\\
                                                                                 & \PWpr and \PZpr triplet & 1 & $\pm$1, 0 & $\qqbar'$, $\qqbar$ & $\qqbar'$, $\qqbar$\\
\hline
\multirow{3}{*}{HVT model B, $g_\mathrm{V} = 3$} & \PWpr singlet & 1 & $\pm$1 & $\qqbar'$ & WZ, WH \\
                                                                                 & \PZpr singlet & 1 & 0 & $\qqbar$ &  WW, ZH \\
                                                                                 & \PWpr and \PZpr triplet & 1 & $\pm$1, 0 & $\qqbar'$, $\qqbar$ & WZ, WH, WW, ZH\\
\hline
RS bulk, $\tilde{k}=0.5$ & $\Gbulk$ & 2 & 0 & gg & WW, ZZ \\
\end{tabular}
}
\label{tab:models}
\end{table*}

\section{Data analyses}
\label{sec:analyses}
\subsection{The CMS detector}

The central feature of the CMS apparatus is a superconducting solenoid of 6\unit{m} internal diameter, providing a magnetic field of 3.8\unit{T}. Within the solenoid volume are a silicon pixel and strip tracker, a lead tungstate crystal electromagnetic calorimeter, and a brass and scintillator hadron calorimeter, each composed of a barrel and two endcap sections. Forward calorimeters extend the pseudorapidity coverage provided by the barrel and endcap detectors. Muons are measured in gas-ionization detectors embedded in the steel flux-return yoke outside the solenoid. A more detailed description of the CMS detector, together with a definition of the coordinate system used and the relevant kinematic variables, can be found in Ref.~\cite{Chatrchyan:2008zzk}.

\subsection{Analysis techniques}

This paper combines searches for heavy resonances over a background spectrum
described by steeply falling distributions of the invariant mass of two reconstructed W, Z, or Higgs bosons in several decay modes.
The $\rm Z\to \ell\ell$ candidates are reconstructed from electron~\cite{Khachatryan:2015hwa} or muon~\cite{Chatrchyan:2012xi} candidates, while $\PW \to \ell\Pgn$ candidates are formed from the combination of electron or muon candidates with missing transverse momentum~\cite{Khachatryan:2014gga},
where the longitudinal momentum of the neutrino is constrained such that the $\ell\Pgn$ invariant mass is equal to the W mass~\cite{Olive:2016xmw}.
The selection criteria for leptons are such that they ensure disjoint datasets for the searches in lepton+jet final states with 0, 1, and 2 leptons.
The contributions from $\PH\to\tau\tau$ candidates are constructed from e and $\mu$ decays of $\tau\to\ell \PAGn_\ell \Pgn_\tau$,
and from $\tau\to\qqbar'\Pgn_\tau$ candidates, in combination with missing transverse momentum.
The $\PW\to\qqbar'$, $\PZ\to\qqbar$, $\PH\to\bbbar$, and $\PH\to\PW\PW\to\qqbar'\qqbar'$ candidates are reconstructed from QCD-evolved jets~\cite{Khachatryan:2014vla}, as described in detail in the following.

Since the W, Z, and Higgs bosons originating from decays of heavy resonances tend to have large Lorentz boosts, their decay products have a small angular separation, requiring special reconstruction techniques.
For highly boosted W, Z, and Higgs bosons decaying to electron, muon, and tau candidates, identification and isolation requirements are formulated such that any other nearby reconstructed lepton is excluded from the computation of quantities used for identification and isolation. This method retains high identification efficiency, while maintaining the same misidentification probability when two leptons are very collimated.

When W, Z, or Higgs bosons decay to quark-antiquark pairs, the showers of hadrons originating from these pairs merge into single large-radius jets that are reconstructed using two jet algorithms~\cite{Cacciari:2011ma}.
The Cambridge--Aachen~\cite{Wobisch:1998wt} and the anti-\kt~\cite{Cacciari:2008gp} algorithms with a distance parameter of 0.8 are used for the 8 and 13\TeV data, respectively, providing comparable jet reconstruction performance.
Jet momenta are corrected for additional pp collisions (pileup) that overlap the event of interest, as specified in Ref.~\cite{CMS:JetCalibration}.
To discriminate against quark and gluon jet background, selections on the pruned jet mass~\cite{Ellis:2009su,Ellis:2009me} and the N-subjettiness ratio $\tau_{2}/\tau_{1}$~\cite{Thaler:2010tr} are applied.
The jet pruning algorithm reclusters the jet constituents, while applying additional requirements to eliminate soft, large-angle QCD radiation that increases the jet mass relative to the initial V or H, quark, or gluon jet mass.
The variable $\tau_{2}/\tau_{1}$ indicates the probability of a jet to be composed of two hard subjets rather than just one hard jet.
A jet is a candidate V jet if its pruned mass, \mJ, is compatible within resolution with the W or Z mass.
The specific selection depends on the analysis channel. For example, the 13\TeV analyses define the window in the range $65 < \mJ < 105\GeV$.
In the 13\TeV data, to further enhance analysis sensitivity to different signal hypotheses, two distinct categories enriched in W or Z bosons are defined through two disjoint ranges in \mJ.
Sensitivity is then further improved in both 8 and \unit{13}{\TeV} data by categorizing events according to the $\tau_{2}/\tau_{1}$ variable into a low purity (LP) and a high purity (HP) category.
Although the HP category dominates the total sensitivity of the analyses, the LP category is retained, since it provides improved sensitivity for high-mass resonances.
The optimal selection criteria for \mJ and $\tau_{2}/\tau_{1}$ depend on signal and background yields and therefore differ across analyses. As a consequence, the efficiencies for identifying W and Z bosons can be different. The total efficiency of the \mJ and $\tau_{2}/\tau_{1}$ HP selection criteria for a jet with \pt of 1\TeV originating from the decay of a heavy resonance ranges from 45\% to 75\%, with a mistagging rate of 2\% to 7\%~\cite{Khachatryan:2014vla,CMS-PAS-JME-16-003}.

A category enriched in Higgs bosons is identified through a pruned-jet mass window around the Higgs boson mass, ensuring a separate selection relative to V jet identification.
For example, the searches in the $\Pgn\Pgn\bbbar$, $\ell\Pgn\bbbar$, and $\ell\ell\bbbar$ final states at 13\TeV~\cite{Khachatryan:2016cfx} define the window in the range $105 < \mJ < 135\GeV$.
In addition, for the $\bbbar$ final state, further discrimination against background is gained by applying a b tagging algorithm~\cite{Chatrchyan:2012jua,CMS:BTV13001,CMS-PAS-BTV-15-001} to the two individual subjets into which the H-jet candidate is split.
The b tagging algorithm discriminates jets originating from b quarks against those originating from lighter quarks or gluons.
To distinguish $\PH\to \PW\PW\to \qqbar'\qqbar'$ jets from background, a technique similar to V jet identification is applied using the $\tau_{4}/\tau_{2}$ N-subjettiness ratio~\cite{Khachatryan:2015bma}. The selection efficiencies for each signal and channel are summarized in Table~\ref{tab:efficiencies}.

\begin{table*}[htb]
  \centering
  \topcaption{Summary of signal efficiencies in analysis channels for 2\TeV resonances in the different models under study.
For analyses that define high-purity (HP) and low-purity (LP) categories, both efficiencies are quoted in the form HP/LP.
Signal efficiencies are given in percent, and include the SM branching fractions of the bosons to the final state in the analysis channel,
effects from detector acceptance, as well as reconstruction and selection efficiencies.
Dashes indicate negligible signal contributions that are not considered in the overall combination.
Channels marked with an asterisk have been reinterpreted for this combination, as described in the text later.}
  \begin{tabular}{lcc cc c{c}@{\hspace*{5pt}}c c}
  \multicolumn{1}{c}{} & \multicolumn{1}{c}{} & \multicolumn{7}{c}{Efficiency [\%]}\\ \cline{3-9}
   & & \multicolumn{4}{c}{HVT} && \multicolumn{2}{c}{RS bulk} \\ \cline{3-6} \cline{8-9}
  Channel & Ref. & \multicolumn{2}{c}{\PWpr} & \multicolumn{2}{c}{\PZpr} && \multicolumn{2}{c}{$\Gbulk$} \\
  & & WZ & WH & WW & ZH && WW & ZZ \\
    \hline
 3$\ell\Pgn$ (8\TeV) & \cite{Khachatryan:2014xja} & 0.6 & \NA & \NA & \NA && \NA & \NA\\
 $\ell\ell \qqbar$ (8\TeV) & \cite{Khachatryan:2014gha} &  *1.1/\NA &  \NA &  \NA &  *0.2/\NA &&  \NA &  3.0/1.0\\
 $\ell \Pgn \qqbar$ (8\TeV) & \cite{Khachatryan:2014gha} &  *4.8/\NA &  \NA &  *9.4/\NA &  \NA &&  10.6/7.1 &  \NA\\
 $\qqbar \qqbar$ (8\TeV) & \cite{Khachatryan:2014hpa} &  5.9/5.5 &  *0.8/0.7 &  *5.7/5.3 & *0.8/0.7 &&  3.8/3.1 &  5.7/4.2\\
    \hline
 $\ell \Pgn \bbbar$(8\TeV) & \cite{Khachatryan:2016yji} &  \NA &  0.9 &  \NA &  \NA &&  \NA &  \NA\\
 $\qqbar \tau\tau$ (8\TeV)  & \cite{Khachatryan:2015ywa} &  \NA &  *1.2 &  \NA &  1.3 &&  \NA &  \NA\\
 $\qqbar\bbbar/6\Pq$ (8\TeV) & \cite{Khachatryan:2015bma} &  \NA &  3.0/1.8 &  \NA &  1.7/1.1 &&  \NA &  \NA\\
    \hline
 $\ell \Pgn \qqbar$ (13\TeV) & \cite{Sirunyan:2016cao} &  10.2 &  1.7 &  19.4 &  \NA &&  18.1 &  \NA\\
 $\qqbar \qqbar$ (13\TeV) & \cite{Sirunyan:2016cao} &  9.7/12.3 &  1.8/2.5 &  8.2/10.6 & 1.9/2.6 &&  8.7/12.4 &  11.0/13.5\\
    \hline
 $\ell\ell\bbbar$ (13\TeV) & \cite{Khachatryan:2016cfx} &  \NA &  \NA &  \NA &  1.5 &&  \NA &  \NA\\
 $\ell\Pgn\bbbar$ (13\TeV) & \cite{Khachatryan:2016cfx} &  \NA &  4.0 &  \NA &  \NA &&  \NA &  \NA\\
 $\Pgn\Pgn\bbbar$ (13\TeV)  & \cite{Khachatryan:2016cfx} &  \NA &  \NA &  \NA &  4.2 &&  \NA &  \NA\\
  \end{tabular}
  \label{tab:efficiencies}
\end{table*}

In all-jet final states~\cite{Khachatryan:2014hpa,Khachatryan:2015bma,Sirunyan:2016cao}, the background expectation is dominated by multijet production, which is estimated through a fit of a signal+background hypothesis to the data, where the background is described by a smoothly falling parametric function. In lepton+jet ($\ell \Pgn \qqbar$, $\ell\ell \qqbar$, $\Pgn \Pgn \bbbar$, $\ell \Pgn \bbbar$, $\ell \ell \bbbar$, and $\qqbar \tau\tau$) final states~\cite{Khachatryan:2014gha,Khachatryan:2016yji,Khachatryan:2015ywa,Sirunyan:2016cao,Khachatryan:2016cfx}, the dominant backgrounds from V+jets production are estimated using data in the sidebands of $\mJ$.
The contamination from WH and ZH resonances decaying into lepton+jet final states in the high sideband defined in the $\ell \Pgn \qqbar$ and $\ell\ell \qqbar$ analyses has been evaluated considering the cross sections excluded by the $\ell \Pgn \bbbar$ and $\ell \ell \bbbar$ searches.
The impact of this contamination on the resulting background estimate is found to be negligible.
In all-lepton final states~\cite{Khachatryan:2014xja}, the dominant background from SM diboson production is estimated using simulated events.

\subsection{Reinterpretations}
In this subsection, we discuss analyses that have been reinterpreted for this paper since not all signal models presented in this combination were considered in the originally published analyses.

In the searches for new heavy resonances decaying into pairs of vector bosons in lepton+jet ($\ell \Pgn \qqbar$ and $\ell\ell \qqbar$) final states~\cite{Khachatryan:2014gha} at $\sqrt{s} = 8\TeV$, 95\% confidence level (CL) exclusion limits are obtained for the production cross section of a bulk graviton.
Using a parametrization for the reconstruction efficiency as a function of W and Z boson kinematics, a reinterpretation is performed in the context of the HVT model described in Section~\ref{subsec:hvt}, which predicts the production of charged and neutral spin-1 resonances decaying preferably to WW and WZ pairs.
This reinterpretation is obtained by rescaling the bulk-graviton signal efficiencies by factors taking into account the different kinematics of W and Z bosons from \PWpr{} and \cPZpr{} production relative to graviton production.
The scale factors are obtained for each value of the sought resonance by means of the tables published in Ref.~\cite{Khachatryan:2014gha}.
Signal shapes are unchanged by the combination process, and the effect of the scaling factor on the signal efficiency takes into account the differences in acceptance for the various signals and masses.
Since the parametrization is restricted to the HP category of the analyses, the LP category is not used for the HVT \PWpr{} and \cPZpr{} interpretations of these channels.
The \mJ window that defines the signal regions of the analysis channels is chosen such that the $\ell \Pgn \qqbar$ channel is sensitive to both the charged and the neutral resonances predicted in the HVT model.
This additional signal efficiency is taken into account in the combination presented in Section~\ref{sec:hvt_results}.

The searches for heavy resonances decaying into pairs of vector bosons in the lepton+jet ($\ell \Pgn \qqbar$ and $\ell\ell \qqbar$)~\cite{Khachatryan:2014gha,Khachatryan:2014gha,Sirunyan:2016cao} and all-jet ($\qqbar \qqbar$)~\cite{Khachatryan:2014hpa,Sirunyan:2016cao} final states at 8 and 13\TeV are also sensitive to the WH and ZH signatures, since a small fraction of jets initiated by Higgs bosons have a pruned jet mass in the W or Z range.
These searches are therefore reinterpreted for WH and ZH signals, to profit from this additional sensitivity.
The efficiencies of these additional signals for the analyses selections are calculated and indicated in Table~\ref{tab:efficiencies} with an asterisk.
This contribution is found to be negligible for the search in the $\ell \Pgn \qqbar$ final state at 8\TeV, as in this analysis events are rejected if the boson jet satisfies b tagging requirements.
The fraction of jets initiated by Z bosons that have a pruned jet mass in the Higgs boson mass range is found to be negligible and therefore this contribution is not taken into account in the combination.

The search for resonances in the $\qqbar \tau\tau$ final state~\cite{Khachatryan:2015bma} is optimized for a \PZpr resonance decaying to a ZH pair.
However, given the large \mJ window ($65 <  \mJ < 105\GeV$) used to tag the $\PZ\to\qqbar$ decays, this analysis channel is also sensitive to the production of the charged spin-1 \PWpr resonance decaying to a WH pair predicted in HVT models.
Similarly, the search in the all-jet final state with 8\TeV data is optimized for the $\PWpr\to\PW\Z$ signal hypothesis, while being sensitive as well to a \PZpr resonance decaying to WW.
This overlap is taken into account in the statistical combination described in Section~\ref{sec:hvt_results}.
For all the other analyses, limits have been previously obtained in the same models as those considered in this letter and a reinterpretation is not needed.

\section{Combination procedure}
\label{sec:combination}
We search for a peak on top of a falling background spectrum by means of a fit to the data.
The likelihood function is constructed using the diboson invariant mass distribution in data,
the background prediction, and the resonant line-shape, to assess the presence of a potential diboson resonance.
We define the likelihood function $\mathcal{L}$ as
\begin{equation}\label{eq:likelihood}
\mathcal{L}(\text{data}~|~\mu \, s(\theta) + b(\theta) ) = \mathcal{P}(\text{data}~|~\mu \, s(\theta) + b(\theta)) \, p(\tilde{\theta}|\theta),
\end{equation}
where ``data'' stands for the observed data; $\theta$ represents the full ensemble of nuisance parameters; $s(\theta)$ and $b(\theta)$ are the expected signal and background yields;
$\mu$ is a scale factor for the signal strength;
$\mathcal{P}(\text{data}~|~\mu \, s(\theta) + b(\theta))$ is the product of Poisson probabilities over all bins of diboson invariant mass distributions in all channels (or over all events for channels with unbinned distributions);
and $p(\tilde{\theta} | \theta)$ is the probability density function for all nuisance parameters to measure a value $\tilde{\theta}$ given its true value $\theta$~\cite{CMS-NOTE-2011-005}.
After maximizing the likelihood function, the best-fit value of $\mu=\sigma_\text{best-fit}/\sigma_\text{theory}$ corresponds therefore to the ratio of the best-fit signal cross section $\sigma_\text{best-fit}$ to the predicted cross section $\sigma_\text{theory}$,
assuming that all branching fractions are as predicted by the relevant signal models.

The treatment of the background in the maximum likelihood fit depends on the analysis channel.
In the $\qqbar \qqbar$, $\qqbar\bbbar$, and $6\Pq$ analyses, the parameters in the background function are left floating in the fit, such that the background prediction is obtained simultaneously with $\mu$, in each hypothesis~\cite{Khachatryan:2014hpa}.
In the remaining analyses ($\ell \Pgn \qqbar$,  $\ell\ell \qqbar$, $\ell \ell \bbbar$, $\ell \Pgn \bbbar$, $\Pgn \Pgn \bbbar$), the background is estimated using sidebands in data, and the uncertainties related to its parametrized distribution are treated as nuisance parameters constrained through Gaussian probability density functions in the fit~\cite{Khachatryan:2014gha}.
The likelihoods from all analysis channels are combined.

The asymptotic approximation~\cite{AsymptCLs} of the $\mathrm{CL_s}$ criterion~\cite{CLs3,CLs1} is used to obtain limits on the signal scale factor $\mu$ that take into account the ratio of the theoretical predictions for the production cross sections at 8 and 13\TeV.

Systematic uncertainties in the signal and background yields are treated as nuisance parameters constrained through log-normal probability density functions.
All such parameters are profiled (refitted as a function of the parameter of interest $\mu$) in the maximization of the likelihood function.
When the likelihoods from different analysis channels are combined, the correlation of systematic effects across those channels is taken into account by treating the uncertainties as fully correlated (associated with the same nuisance parameter) or fully uncorrelated (associated with different nuisance parameters).
Table~\ref{tab:correlations} summarizes which uncertainties are treated as correlated among 8 and 13\TeV analyses, e and $\mu$ channels, HP and LP categories,
and mass categories enriched in W, Z, and Higgs bosons in the combination.
Additional categorization within individual analyses is described in their corresponding papers.
The nuisance parameters treated as correlated between 8 and 13\TeV analyses are those related to the parton distribution functions (PDFs) and the choice of the factorization ($\mu_\mathrm{f}$) and renormalization ($\mu_\mathrm{r}$) scales used to estimate the signal cross sections.
The signal cross sections and their associated uncertainties are reevaluated for this combination at both 8 and 13\TeV, estimating thereby their full impact on the expected signal yield rather than just the impact on the signal acceptance.
The PDF uncertainties are evaluated using the NNPDF 3.0~\cite{Ball:2011mu} PDFs.
The uncertainty related to the choice of $\mu_\mathrm{f}$ and $\mu_\mathrm{r}$ scales is evaluated following~\cite{Cacciari:2003fi,Catani:2003zt} by changing the default choice of scales in six combinations of
$(\mu_\mathrm{f}$, $\mu_\mathrm{r})$ by factors of $(0.5, 0.5)$, $(0.5, 1)$, $(1,0.5)$, $(2, 2)$, $(2, 1)$, and $(1, 2)$.
The experimental uncertainties are all treated as uncorrelated between 8 and 13\TeV analyses.
The case where the most important uncertainties are treated as fully correlated among 8 and 13\TeV analyses has been studied and found to have negligible impact on the results.
After the combined fit, no nuisance parameter was found to differ significantly from its expectation and from the fit result in individual analyses.

\begin{table*}[htb]
  \centering
  \topcaption{Correlation across analyses of systematic uncertainties in the signal prediction affecting the event yield in the signal region and the reconstructed diboson invariant mass distribution. A ``yes'' signifies 100\% correlation, and ``no'' means uncorrelated.}
  \cmsTableResize{
  \begin{tabular}{ll*{4}{c}}
    Source & Quantity & 8 and 13\TeV & e and $\mu$ & HP and LP & W-, Z-, and H-enriched \\
    \hline
    Lepton trigger & yield & no & no & yes & yes \\
    Lepton identification & yield & no & no & yes & yes \\
    Lepton momentum scale & yield, shape & no & no & yes & yes \\
    Jet energy scale & yield, shape & no & yes & yes & yes \\
    Jet energy resolution & yield, shape & no & yes & yes & yes \\
    Jet mass scale & yield & no & yes & yes & yes \\
    Jet mass resolution & yield & no & yes & yes & yes \\
    b tagging & yield & no & yes & yes & yes \\
    W tagging \nsubj{} (HP/LP) & yield & no & yes & yes & yes \\
    Integrated luminosity & yield & no & yes & yes & yes \\
    Pileup & yield & no & yes & yes & yes \\
    PDF & yield & yes & yes & yes & yes \\
    $\mu_\mathrm{f}$ and $\mu_\mathrm{r}$ scales & yield & yes & yes & yes & yes \\
  \end{tabular}
  }
  \label{tab:correlations}
\end{table*}

\section{Results}
\label{sec:results}
We evaluate the combined significance of the 8 and 13\TeV CMS searches for all signal hypotheses.
The ATLAS Collaboration reported an excess in the all-jet VV~$\to\qqbar\qqbar$ search, corresponding to a local significance of 3.4 standard deviations (s.d.) for a \PWpr resonance with a mass of 2\TeV~\cite{Aad:2015owa}. Similarly, the CMS experiment reported a local deviation of 2.2 s.d. in the lepton+jet $\PW\PH\to\ell\Pgn\bbbar$ search for a \PWpr resonance with a mass of 1.8\TeV~\cite{Khachatryan:2016yji}.
The present combination does not confirm these small excesses (within the context of the models considered), as the highest combined significance in the mass range of the reported excesses is found to be for a \PWpr resonance at 1.8\TeV with a local significance of 0.8 standard deviations.

In the following, we present for each channel 95\% CL exclusion limits on the signal strength $\mu$ in Eq.~\ref{eq:likelihood},
expressed as the exclusion limit on the ratio $\sigma_{95\%}/\sigma_\text{theory}$ of the signal cross section to the predicted cross section,
assuming that all branching fractions are as predicted by the relevant signal models.

\subsection{Limits on \PWpr and \PZpr singlets}\label{sec:wp_results}

\begin{figure*}[htbp]
\centering
\includegraphics[width=0.48\textwidth]{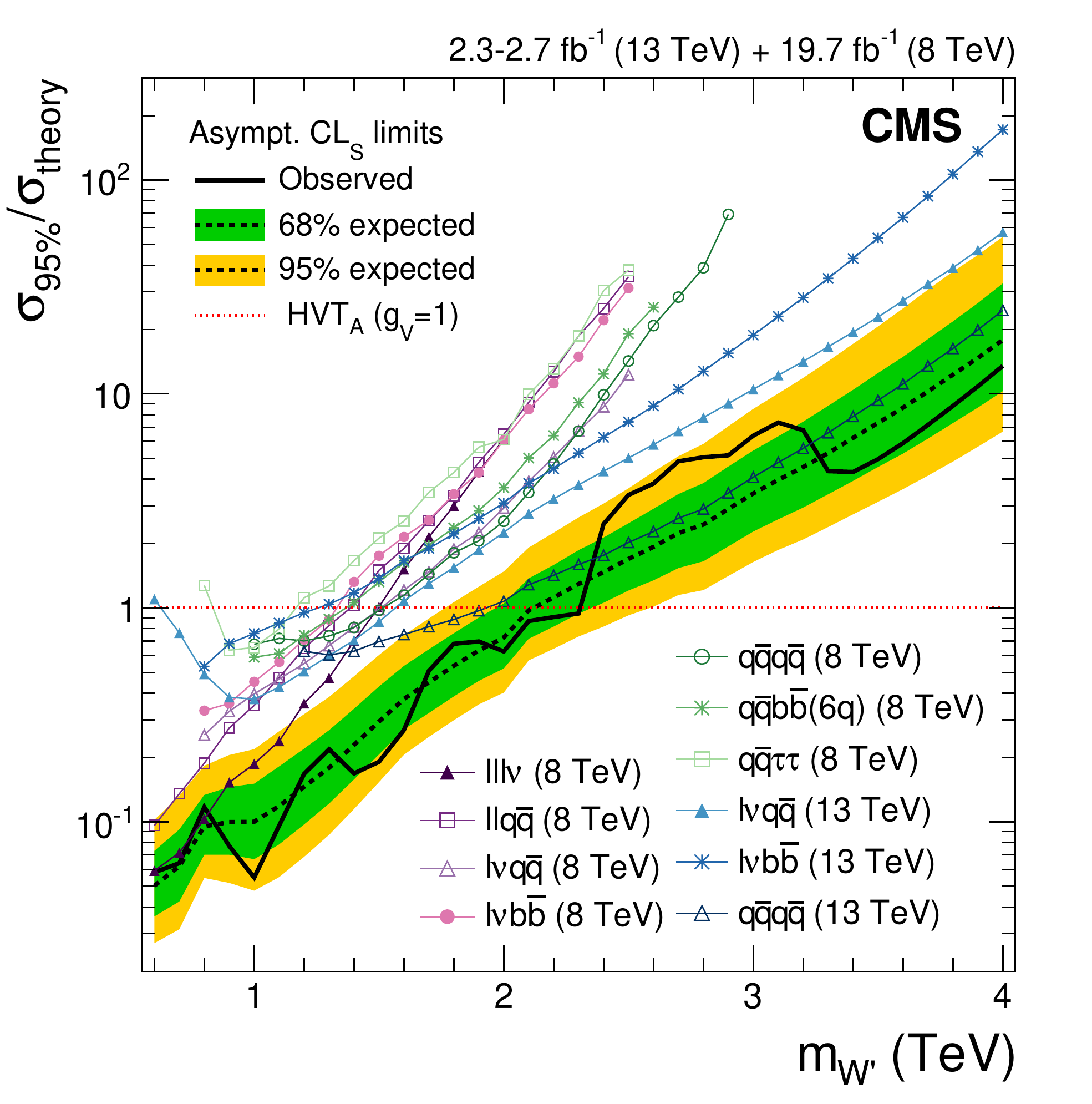}
\includegraphics[width=0.48\textwidth]{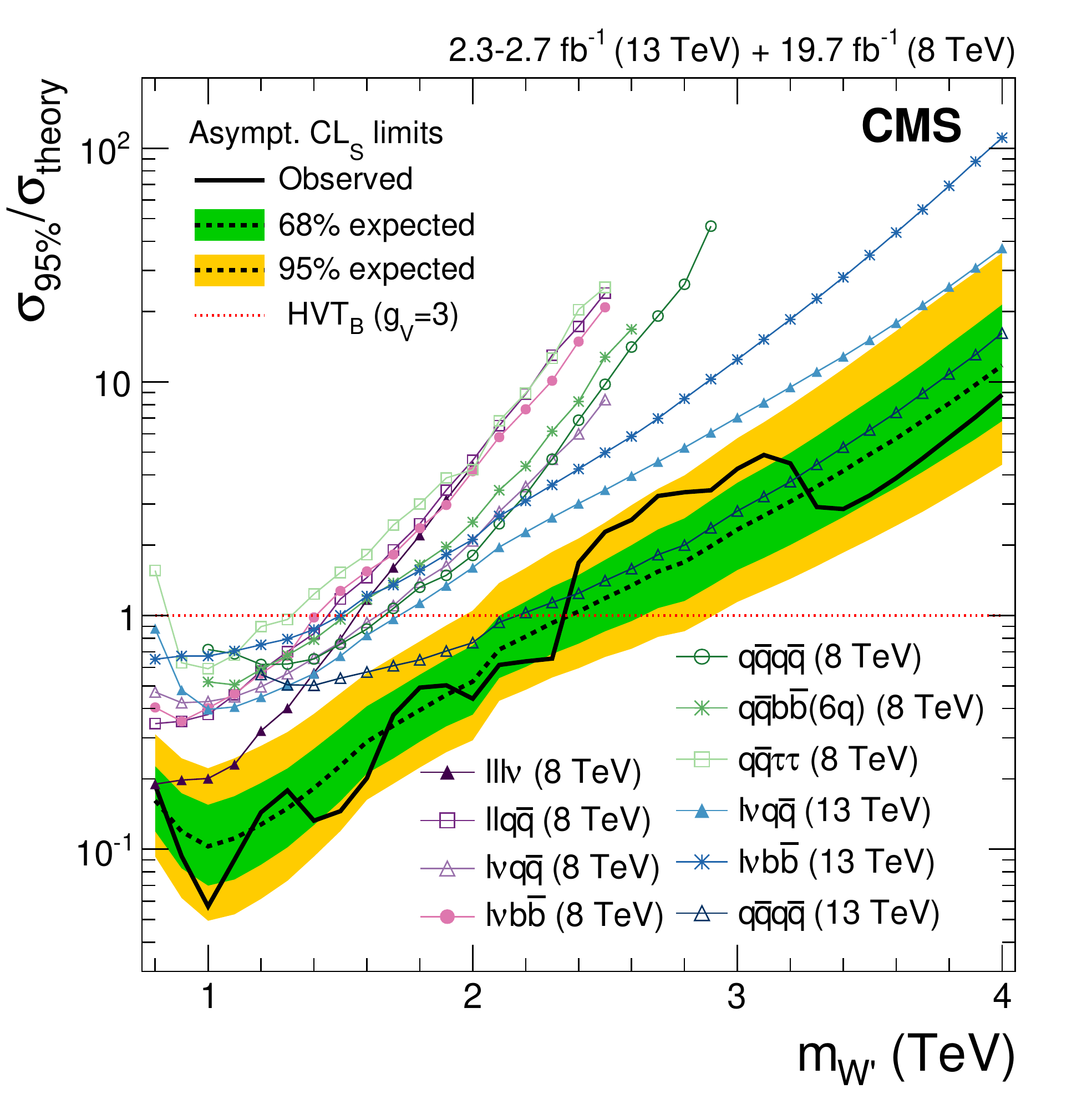}\\
\includegraphics[width=0.48\textwidth]{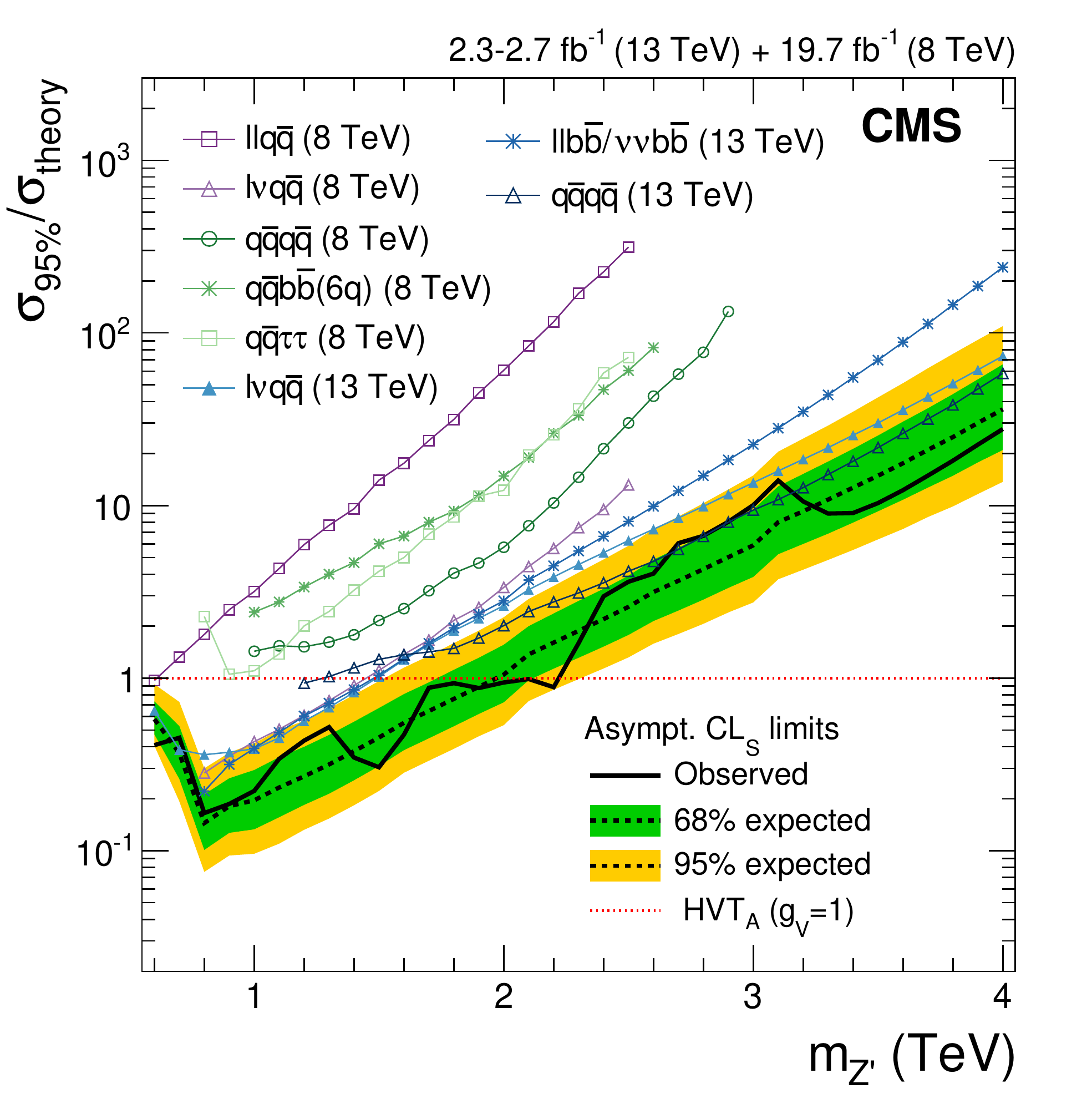}
\includegraphics[width=0.48\textwidth]{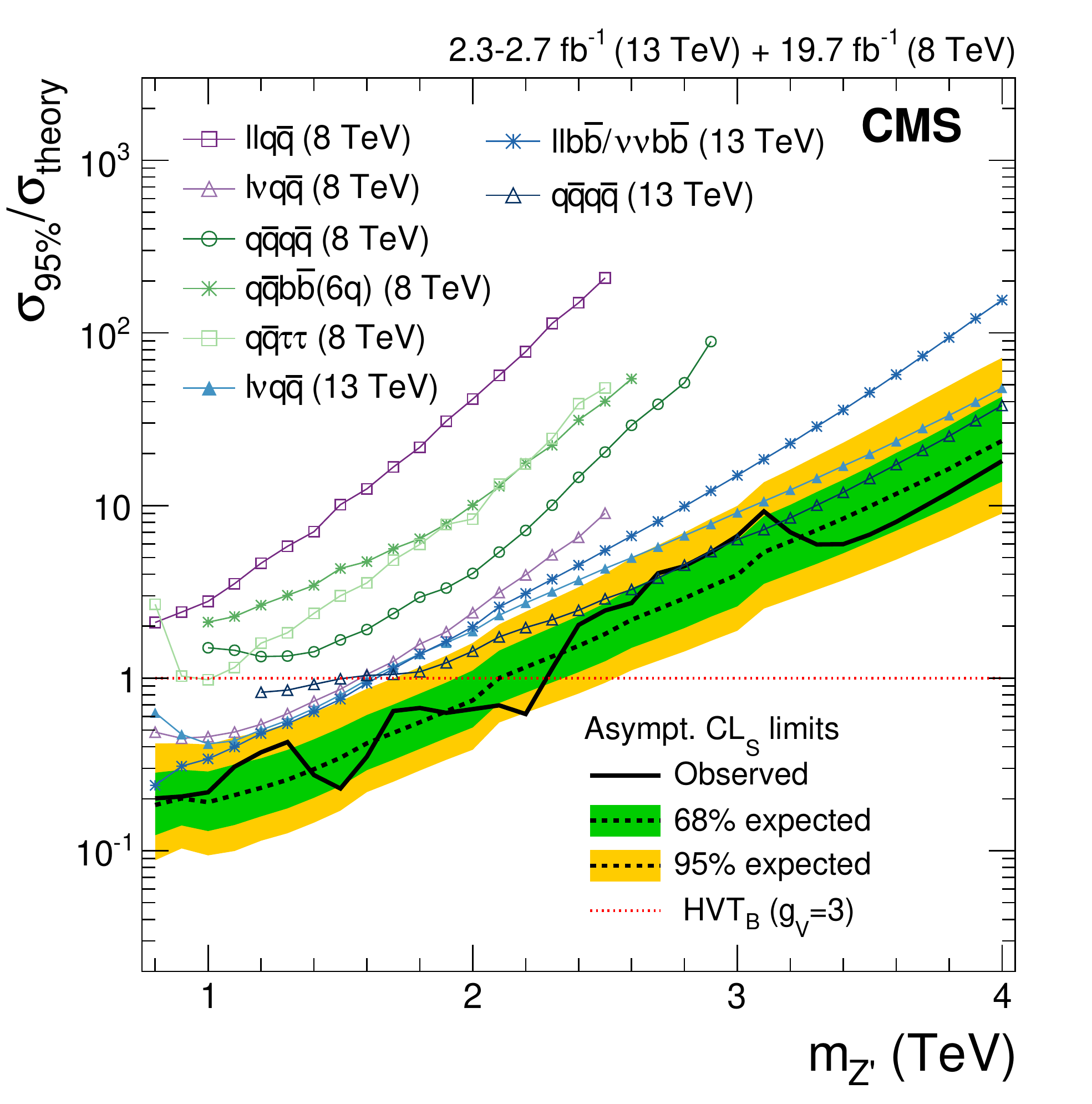}
\caption{Exclusion limits at 95\% CL for HVT models A (left) and B (right) on the signal strengths for the singlets $\PWpr\to\PW\PZ$ and $\PW\PH$ (upper),
and $\PZpr\to\PW\PW$ and $\PZ\PH$ (lower) as a function of the resonance mass, obtained by combining the 8 and 13\TeV analyses.
The signal strength is expressed as the ratio $\sigma_{95\%}/\sigma_\text{theory}$ of the signal cross section to the predicted cross section,
assuming that all branching fractions are as predicted by the relevant signal models.
The curves with symbols refer to the expected limits obtained by the analyses that are inputs to the combinations. The thick solid (dashed) line represents the combined observed (expected) limits.}
\label{fig:wpall_138TeV}
\end{figure*}

Figure~\ref{fig:wpall_138TeV} (upper) shows a comparison and combination of results obtained in the 8 and 13\TeV searches for a \PWpr singlet resonance in HVT models A and B.
The 95\% CL exclusion limits on the signal strengths are given for the mass ranges $0.6 < m_{\PWpr} < 4.0\TeV$ for model A and $0.8 < m_{\PWpr} < 4.0\TeV$ for model B.
Table~\ref{tab:HVTlimits} summarizes the lower limits on the resonance masses.
Below mass values of $\approx1.4\TeV$, the most sensitive channel is the 3$\ell\Pgn$ final state at 8\TeV.
At higher masses, the $\qqbar \qqbar$ search at 13\TeV dominates the sensitivity.
The overall sensitivity benefits from the combination for resonance masses up to $\approx$2\TeV,
lowering the exclusion limit on the cross section by up to a factor of $\approx3$ relative to the most sensitive single channel,
as several channels of similar sensitivity are combined in this mass range.
Above resonance masses of 2\TeV, the 8\TeV analyses do not have significant sensitivity compared to the $\qqbar \qqbar$ search at 13\TeV.

\begin{table*}[htb]
  \centering
  \topcaption{Lower limits at 95\% CL on the resonance masses in HVT models A and B. The 68\% quantiles defined as the intervals containing the central 68\% of the distribution of limits expected under the background-only hypothesis are also reported.}
  \begin{tabular}{lccc}
   Model & Observed limit [\TeVns{}] & Expected limit [\TeVns{}] & 68\% quantile\\
    \hline
    Singlet \PWpr (model A)              & 2.3 & 2.1 & [1.9,2.3] \\
    Singlet \PZpr (model A)              & 2.2 & 2.0 & [1.8,2.2] \\
    Triplet \PWpr and \PZpr (model A)    & 2.4 & 2.4 & [2.1,2.7] \\
    \hline
    Singlet \PWpr (model B)              & 2.3 & 2.4 & [2.1,2.7]\\
    Singlet \PZpr (model B)              & 2.3 & 2.1 & [1.9,2.3] \\
    Triplet \PWpr and \PZpr (model B)    & 2.4 & 2.6 & [2.3,2.9] \\
  \end{tabular}
  \label{tab:HVTlimits}
\end{table*}

Figure~\ref{fig:wpall_138TeV} (lower) shows the analogous results for a \PZpr singlet resonance for final states of WW and ZH in the HVT models A and B.
The $\ell \Pgn \qqbar$ channel at 8\TeV and the $\qqbar \qqbar$, $\ell \Pgn \qqbar$, $\ell\ell\bbbar$, and $\Pgn\Pgn\bbbar$ channels at 13\TeV dominate the sensitivity over the whole range, with 8 and 13\TeV analyses giving almost equal contributions for masses below 2\TeV. Above this value, the sensitivity arises mainly from the 13\TeV data.
As in the \PWpr analyses, the mass limit is not affected by the combination compared to what is obtained from the 13\TeV searches.

\subsection{Limits on the heavy vector triplet \texorpdfstring{$\Vprime$}{V'}}\label{sec:hvt_results}

\begin{figure*}[htbp]
\centering
\includegraphics[width=0.48\textwidth]{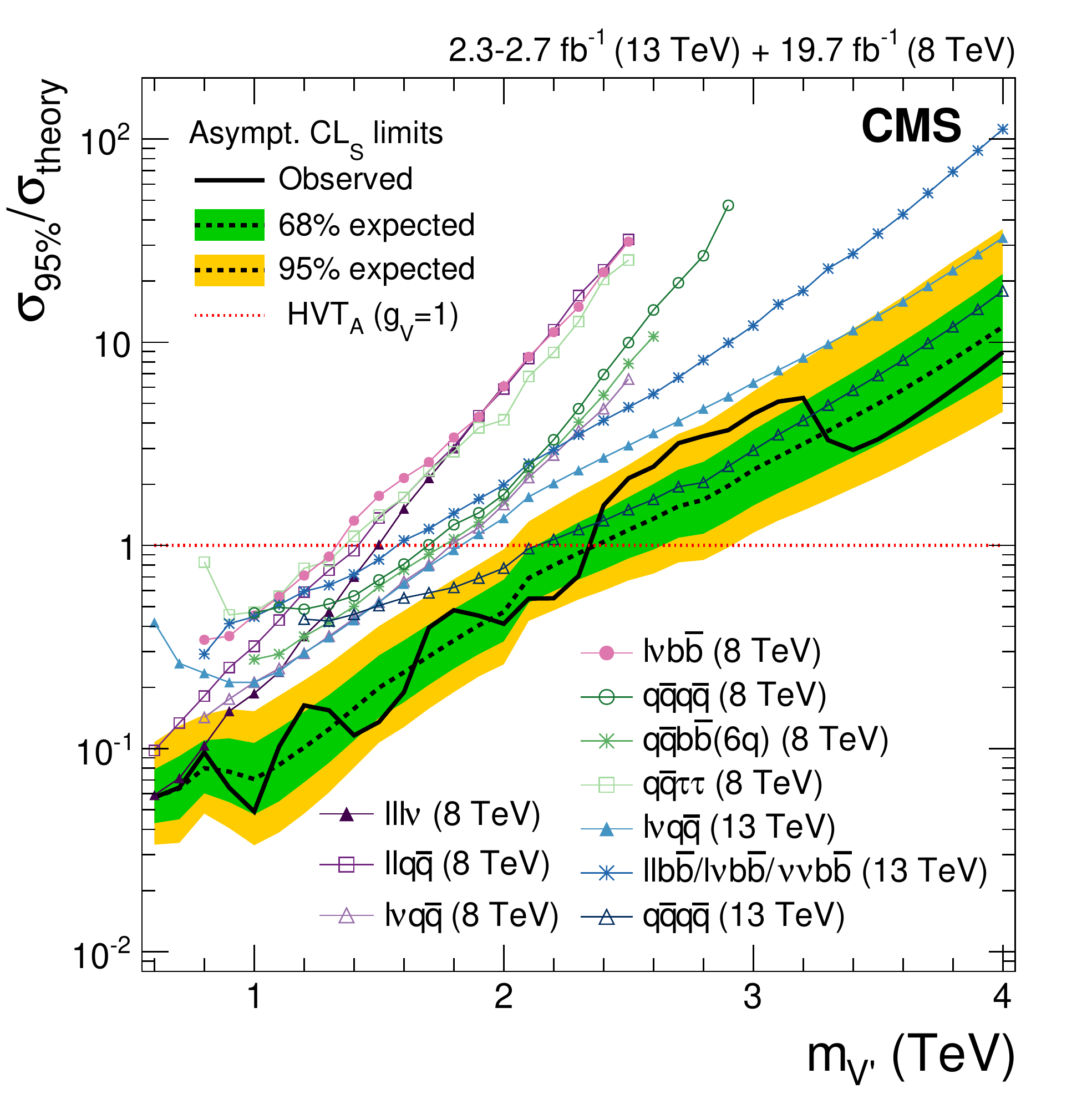}
\includegraphics[width=0.48\textwidth]{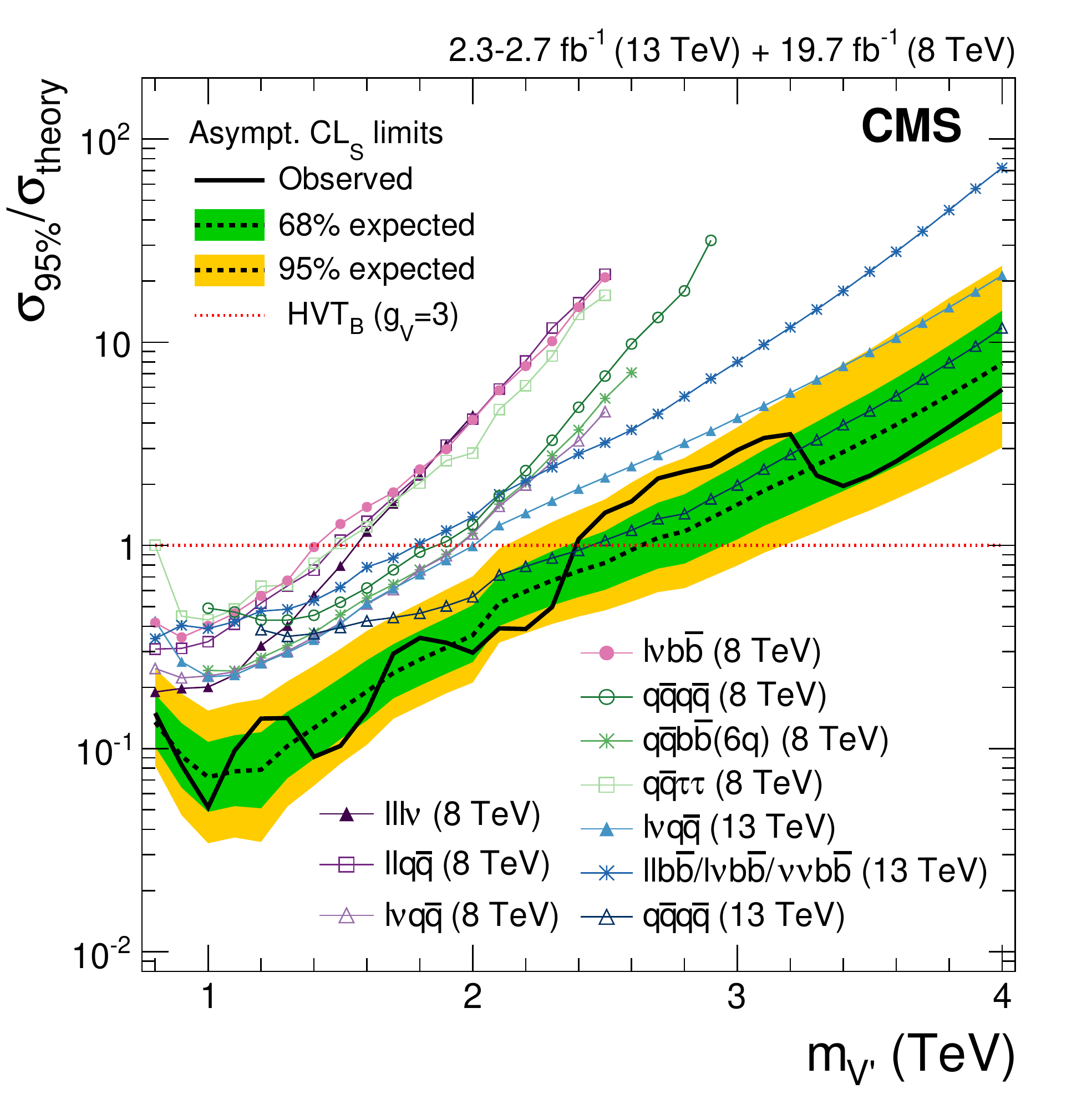}
\includegraphics[width=0.48\textwidth]{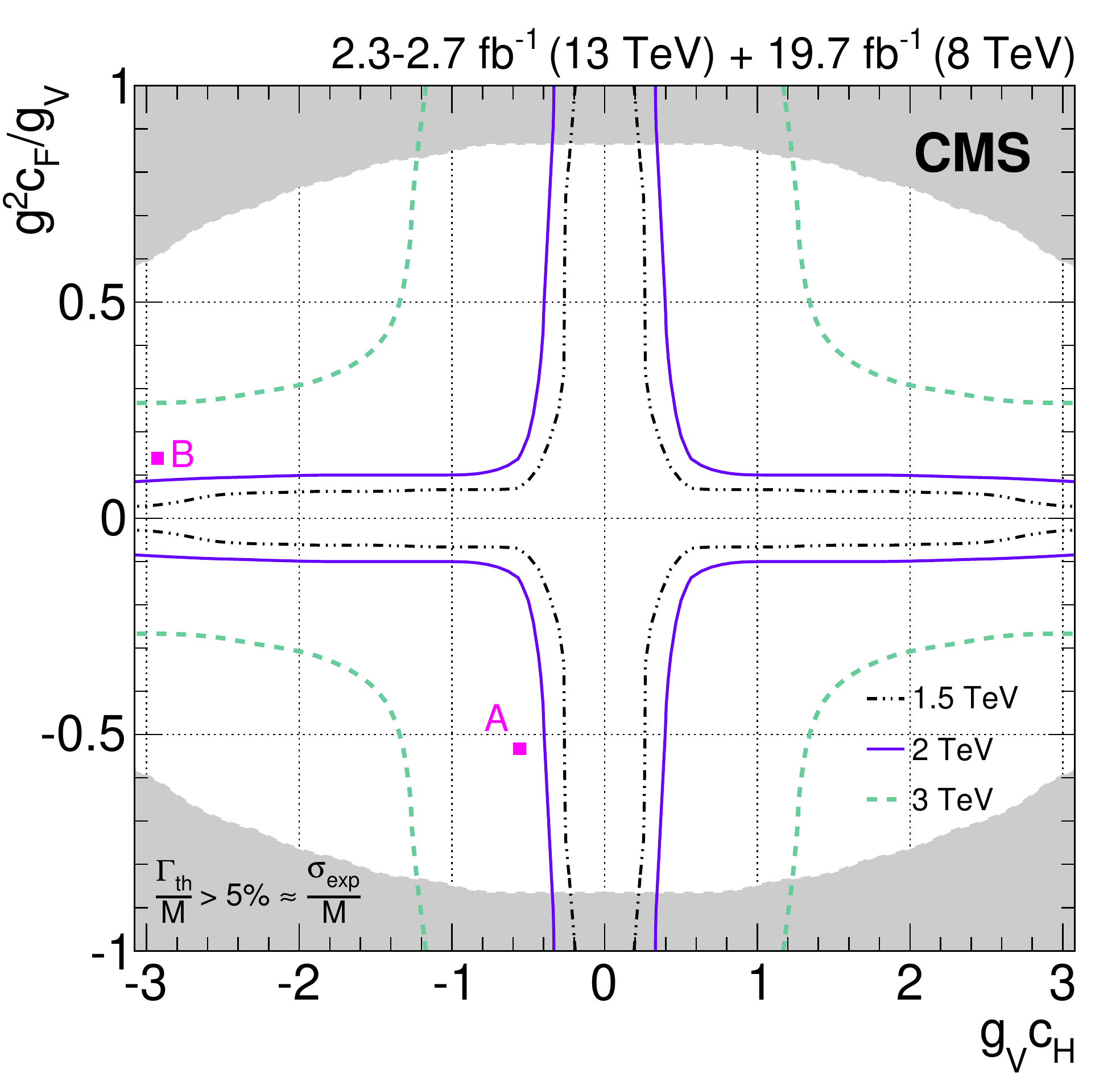}
\caption{
Exclusion limits at 95\% CL on the signal strengths in HVT models A (upper left) and B (upper right) for the triplet $\Vprime$, as a function of the resonance mass, obtained by combining the 8 and 13\TeV diboson searches.
The signal strength is expressed as the ratio $\sigma_{95\%}/\sigma_\text{theory}$ of the signal cross section to the predicted cross section,
assuming that all branching fractions are as predicted by the relevant signal models.
In the upper plots, the curves with symbols refer to the expected limits obtained by the analyses that are inputs to the combinations. The thick solid (dashed) line represents the combined observed (expected) limits.
In the lower plot, exclusion regions in the plane of the HVT-model couplings ($g_\mathrm{V}c_{\PH}$, $g^2c_\mathrm{F}/g_\mathrm{V}$ ) for three resonance masses of 1.5, 2.0, and 3.0\TeV, where $g$ denotes the weak gauge coupling. The points A and B of the benchmark models used in the analysis are also shown.
The boundaries of the regions excluded in this search are indicated by the solid, dashed, and dashed-dotted lines.
The areas indicated by the solid shading correspond to regions where the resonance width is predicted to be more than 5\% of the resonance mass, in which the narrow-resonance assumption is not satisfied.
}
\label{fig:hvtall_138TeV}
\end{figure*}

Figure~\ref{fig:hvtall_138TeV} (upper) shows the comparison and combination of the results obtained in the 8 and 13\TeV searches for resonances in a heavy vector triplet.
The lower limits on the resonance masses for HVT models A and B are quoted in Table~\ref{tab:HVTlimits}. As for the \PWpr and \PZpr cases, the observed mass limit of 2.4\TeV for both models obtained combining the 8 and 13\TeV searches is dominated essentially by the 13\TeV analyses alone.

Figure~\ref{fig:hvtall_138TeV} (lower) displays a scan of the coupling parameters and the corresponding observed 95\% CL exclusion contours in the HVT models from the combination of the 8 and 13\TeV analyses. The parameters are defined as $g_\mathrm{V}c_{\PH}$ and $g^2c_\mathrm{F}/g_\mathrm{V}$ in terms of the coupling strengths of the new resonance to the H and V, and to fermions, respectively, given in Section~\ref{subsec:hvt}. The range is limited by the assumption that the resonance sought is narrow. The shaded area represents the region where the theoretical width is larger than the experimental resolution of the searches, and therefore where the narrow-resonance assumption is not satisfied. This contour is defined by a predicted resonance width, relative to its mass, of 5\%, corresponding to the best detector resolution of the searches.

\subsection{Limits on the bulk graviton}

\begin{figure}[htbp]
\centering
\includegraphics[width=0.48\textwidth]{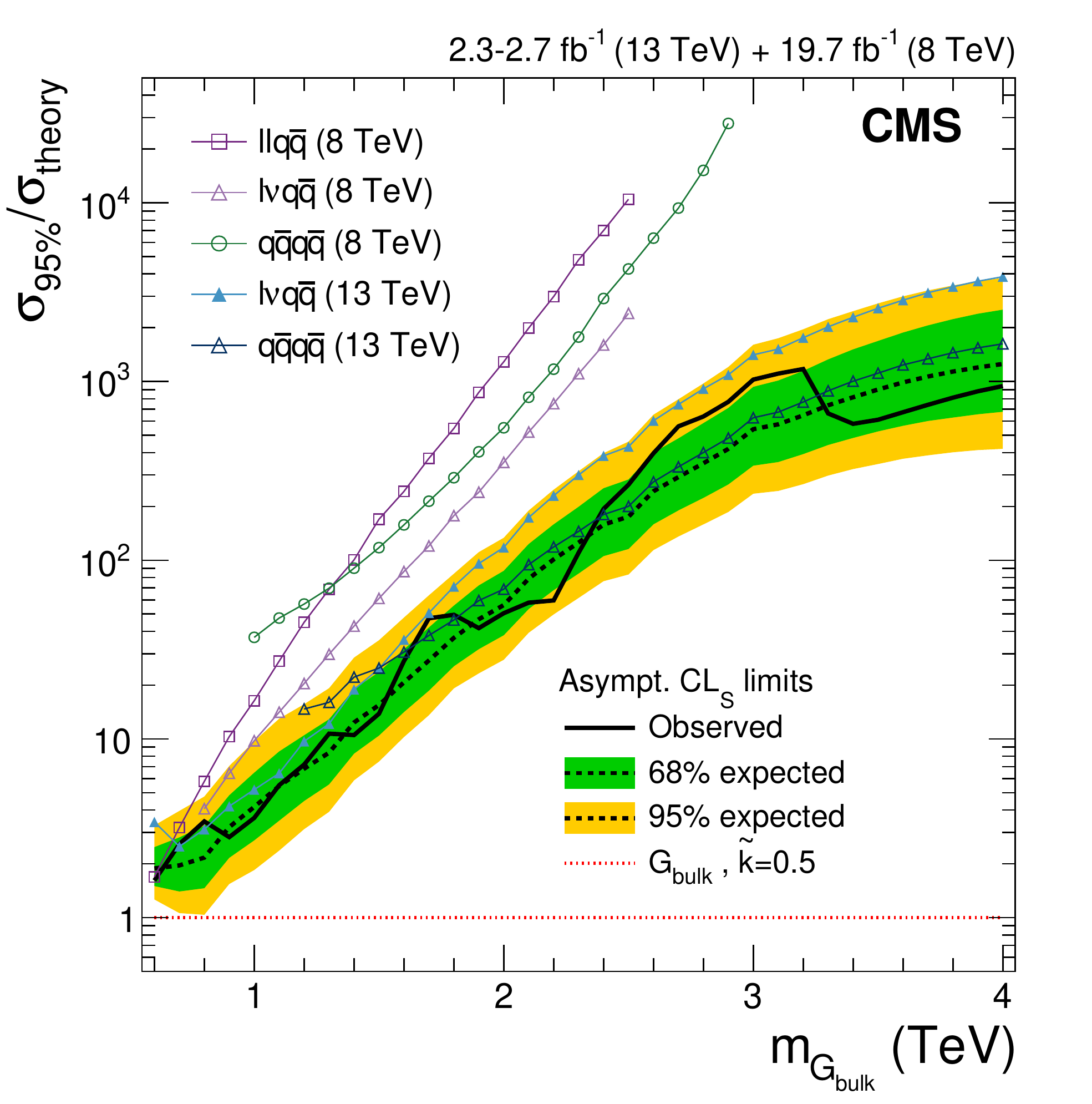}
\caption{
Exclusion limits at 95\% CL on the signal strength in the bulk graviton model with $\tilde{k} = 0.5$, as a function of the resonance mass, obtained by combining the 8 and 13\TeV diboson searches.
The signal strength is expressed as the ratio $\sigma_{95\%}/\sigma_\text{theory}$ of the signal cross section to the predicted cross section,
assuming that all branching fractions are as predicted by the relevant signal models.
The curves with symbols refer to the expected limits obtained by the analyses that are inputs to the combination. The thick solid (dashed) line represents the combined observed (expected) limits.}
\label{fig:bulkgall_138TeV}
\end{figure}

Figure~\ref{fig:bulkgall_138TeV} shows a comparison and combination of results obtained in the 8 and 13\TeV VV searches in the bulk graviton model with $\tilde{k} = 0.5$.
The sensitivity arises mainly from the 13\TeV $\qqbar \qqbar$ and $\ell \Pgn \qqbar$ channels.
The 13\TeV searches supersede the 8\TeV combination down to masses of 0.7\TeV, since in this model, the signal is produced via gluon-gluon fusion, in contrast to the qq annihilation process responsible for the production of HVT resonances.
The combination yields the most stringent limits to date on signal strengths for narrow bulk graviton resonances ($\tilde{k} = 0.5$) in the mass range from 0.6 to 4.0\TeV.

\section{Summary}
\label{sec:conclusions}
A statistical combination of searches for massive narrow resonances decaying to WW, ZZ, WZ, WH, and ZH boson pairs in the mass range 0.6--4.0\TeV has been presented.
The searches are based on proton-proton collision data collected by the CMS experiment at centre-of-mass energies of 8 and 13\TeV, corresponding to integrated luminosities of 19.7 and up to \unit{2.7}{\fbinv}, respectively. The results of the searches and of the combination are interpreted in the context of heavy vector singlet and triplet models predicting \PWpr and \PZpr bosons decaying to WZ, WH, WW, and ZH, and a model with a bulk graviton that decays into WW and ZZ.
The small excesses observed with 8\TeV data by the ATLAS and CMS experiments~\cite{Aad:2015owa,Khachatryan:2016yji} at 1.8--2.0\TeV are not confirmed by the analyses performed with 13\TeV data.
This is the first combined search for WW, WZ, WH, and ZH resonances and yields 95\% confidence level lower limits in the heavy vector triplet model B on the masses of \PWpr and \PZpr singlets at 2.3\TeV, and on a heavy vector triplet at 2.4\TeV.
The limits on the production cross section of a narrow bulk graviton resonance with the curvature scale of the warped extra dimension $\tilde{k}=0.5$, in the mass range of 0.6 to 4.0\TeV, are the most stringent published to date.
The statistical combination of VV and VH resonance searches in several distinct final states was found to yield a significant gain in sensitivity and therefore represents a powerful tool for future resonance searches with the large expected diboson event data sample at the LHC.

\begin{acknowledgments}
We congratulate our colleagues in the CERN accelerator departments for the excellent performance of the LHC and thank the technical and administrative staffs at CERN and at other CMS institutes for their contributions to the success of the CMS effort. In addition, we gratefully acknowledge the computing centers and personnel of the Worldwide LHC Computing Grid for delivering so effectively the computing infrastructure essential to our analyses. Finally, we acknowledge the enduring support for the construction and operation of the LHC and the CMS detector provided by the following funding agencies: BMWFW and FWF (Austria); FNRS and FWO (Belgium); CNPq, CAPES, FAPERJ, and FAPESP (Brazil); MES (Bulgaria); CERN; CAS, MoST, and NSFC (China); COLCIENCIAS (Colombia); MSES and CSF (Croatia); RPF (Cyprus); SENESCYT (Ecuador); MoER, ERC IUT, and ERDF (Estonia); Academy of Finland, MEC, and HIP (Finland); CEA and CNRS/IN2P3 (France); BMBF, DFG, and HGF (Germany); GSRT (Greece); OTKA and NIH (Hungary); DAE and DST (India); IPM (Iran); SFI (Ireland); INFN (Italy); MSIP and NRF (Republic of Korea); LAS (Lithuania); MOE and UM (Malaysia); BUAP, CINVESTAV, CONACYT, LNS, SEP, and UASLP-FAI (Mexico); MBIE (New Zealand); PAEC (Pakistan); MSHE and NSC (Poland); FCT (Portugal); JINR (Dubna); MON, RosAtom, RAS, RFBR and RAEP (Russia); MESTD (Serbia); SEIDI, CPAN, PCTI and FEDER (Spain); Swiss Funding Agencies (Switzerland); MST (Taipei); ThEPCenter, IPST, STAR, and NSTDA (Thailand); TUBITAK and TAEK (Turkey); NASU and SFFR (Ukraine); STFC (United Kingdom); DOE and NSF (USA).

\hyphenation{Rachada-pisek} Individuals have received support from the Marie-Curie program and the European Research Council and Horizon 2020 Grant, contract No. 675440 (European Union); the Leventis Foundation; the A. P. Sloan Foundation; the Alexander von Humboldt Foundation; the Belgian Federal Science Policy Office; the Fonds pour la Formation \`a la Recherche dans l'Industrie et dans l'Agriculture (FRIA-Belgium); the Agentschap voor Innovatie door Wetenschap en Technologie (IWT-Belgium); the Ministry of Education, Youth and Sports (MEYS) of the Czech Republic; the Council of Science and Industrial Research, India; the HOMING PLUS program of the Foundation for Polish Science, cofinanced from European Union, Regional Development Fund, the Mobility Plus program of the Ministry of Science and Higher Education, the National Science Center (Poland), contracts Harmonia 2014/14/M/ST2/00428, Opus 2014/13/B/ST2/02543, 2014/15/B/ST2/03998, and 2015/19/B/ST2/02861, Sonata-bis 2012/07/E/ST2/01406; the National Priorities Research Program by Qatar National Research Fund; the Programa Clar\'in-COFUND del Principado de Asturias; the Thalis and Aristeia programs cofinanced by EU-ESF and the Greek NSRF; the Rachadapisek Sompot Fund for Postdoctoral Fellowship, Chulalongkorn University and the Chulalongkorn Academic into Its 2nd Century Project Advancement Project (Thailand); and the Welch Foundation, contract C-1845.
\end{acknowledgments}
\bibliography{auto_generated}
\appendix
\section{Signal cross section tables}\label{app:xsec}

\begin{table*}[htbp]
  \centering
  \topcaption{Signal cross sections in units of fb at 8\TeV center-of-mass energy.
HVT model A and model B cross sections are quoted in the form $\sigma_{\text{Model A}}/\sigma_{\text{Model B}}$.
}
  \begin{tabular}{*{5}{c}{c}@{\hspace*{5pt}}cc}
  \multicolumn{1}{c}{} & \multicolumn{7}{c}{Cross section at 8\TeV [fb]} \\ \cline{2-8}
  \multicolumn{1}{c}{} & \multicolumn{4}{c}{HVT A/B} && \multicolumn{2}{c}{RS bulk} \\ \cline{2-5}\cline{7-8}
  Mass & \multicolumn{2}{c}{\PWpr} & \multicolumn{2}{c}{\PZpr} && \multicolumn{2}{c}{$\Gbulk$} \\
  \multicolumn{1}{c}{[\TeVns{}]} & WZ & WH & WW & ZH && WW & ZZ \\
    \hline
  0.6 & 1786/\NA &  1377/\NA &  874/\NA &  746/\NA &&  80.7 &  42.4\\
  0.8 & 483/262 &  413/337 &  235/131 &  213/180 &&  12.3 &  6.32\\
  1.0 & 168/155 &  151/171 &  80.0/74.6 &  74.9/85.6 &&  2.75 &  1.41\\
  1.5 & 19.4/24.8 &  18.4/25.5 &  8.85/11.4 &  8.58/11.9 &&  0.142 &  0.0719\\
  2.0 & 2.98/4.19 &  2.89/4.25 &  1.34/1.89 &  1.31/1.93 &&  0.0126 &  0.00627\\
  2.5 & 0.494/0.725 &  0.485/0.731 &  0.227/0.333 &  0.224/0.338 &&  0.00140 &  0.000709\\
  3.0 & 0.0801/0.120 &  0.0791/0.121 &  0.0395/0.0594 &  0.0392/0.0600 &&  \NA&  \NA\\
  \end{tabular}
  \label{tab:xsec8}
\end{table*}

\begin{table*}[htbp]
  \centering
  \topcaption{Signal cross sections in units of fb at 13\TeV center-of-mass energy.
HVT model A and model B cross sections are quoted in the form $\sigma_{\text{Model A}}/\sigma_{\text{Model B}}$.
}
  \begin{tabular}{*{5}{c}{c}@{\hspace*{5pt}}cc}
  \multicolumn{1}{c}{} & \multicolumn{7}{c}{Cross section at 13\TeV [fb]} \\ \cline{2-8}
  \multicolumn{1}{c}{} & \multicolumn{4}{c}{HVT A/B} && \multicolumn{2}{c}{RS bulk} \\ \cline{2-5}\cline{7-8}
  Mass & \multicolumn{2}{c}{\PWpr} & \multicolumn{2}{c}{\PZpr} && \multicolumn{2}{c}{$\Gbulk$} \\
  \multicolumn{1}{c}{[\TeVns{}]} & WZ & WH & WW & ZH && WW & ZZ \\
    \hline
  0.6 & 4170/\NA &  3215/\NA &  2097/\NA &  1789/\NA &&  406.8 &  203.4\\
  0.8 & 1258/680 &  1074/878 &  635/354 &  576/485 &&  76.1 &  38.0\\
  1.0 & 492/464 &  443/501 &  247/229 &  231/264 &&  20.5 &  10.2\\
  1.5 & 81.7/105 &  77.8/108 &  39.8/51.1 &  38.6/53.6 &&  1.80 &  0.901\\
  2.0 & 19.8/27.9 &  19.2/28.3 &  9.32/13.1 &  9.16/13.5 &&  0.240 &  0.120\\
  2.5 & 5.70/8.37 &  5.60/8.44 &  2.61/3.84 &  2.58/3.90 &&  0.0449 &  0.0224\\
  3.0 & 1.79/2.68 &  1.77/2.70 &  0.808/1.21 &  0.801/1.23 &&  0.00982 &  0.00491\\
  3.5 & 0.584/0.888 &  0.579/0.891 &  0.264/0.402 &  0.262/0.405 &&  0.00420 &  0.00210\\
  4.0 & 0.192/0.296 &  0.191/0.296 &  0.0887/0.136 &  0.0883/0.137 &&  0.00244 &  0.00122\\
  \end{tabular}
  \label{tab:xsec13}
\end{table*}

\cleardoublepage \section{The CMS Collaboration \label{app:collab}}\begin{sloppypar}\hyphenpenalty=5000\widowpenalty=500\clubpenalty=5000\textbf{Yerevan Physics Institute,  Yerevan,  Armenia}\\*[0pt]
A.M.~Sirunyan, A.~Tumasyan
\vskip\cmsinstskip
\textbf{Institut f\"{u}r Hochenergiephysik,  Wien,  Austria}\\*[0pt]
W.~Adam, E.~Asilar, T.~Bergauer, J.~Brandstetter, E.~Brondolin, M.~Dragicevic, J.~Er\"{o}, M.~Flechl, M.~Friedl, R.~Fr\"{u}hwirth\cmsAuthorMark{1}, V.M.~Ghete, C.~Hartl, N.~H\"{o}rmann, J.~Hrubec, M.~Jeitler\cmsAuthorMark{1}, A.~K\"{o}nig, I.~Kr\"{a}tschmer, D.~Liko, T.~Matsushita, I.~Mikulec, D.~Rabady, N.~Rad, H.~Rohringer, J.~Schieck\cmsAuthorMark{1}, J.~Strauss, W.~Waltenberger, C.-E.~Wulz\cmsAuthorMark{1}
% ...
\vskip\cmsinstskip
1:~~Also at Vienna University of Technology, Vienna, Austria\\
% ...

\end{sloppypar}
\end{document}